\documentclass[12pt]{iopart}

\begin{document}

\title{A theoretical model for single molecule incoherent scanning tunneling spectroscopy}

\author{H Raza}

\address{NSF Network for Computational Nanotechnology and School of Electrical and Computer Engineering Purdue University West Lafayette Indiana 47907 USA \\School of Electrical and Computer Engineering Cornell University Ithaca New York 14853 USA}
\ead{hr89@cornell.edu}
\begin{abstract}
Single molecule scanning tunneling spectroscopy (STS), with dephasing due to elastic and inelastic scattering, is of some current interest. Motivated by this, we report an extended H\"uckel theory (EHT) based mean-field Non-equilibrium Green's function (NEGF) transport model with electron-phonon scattering treated within the self-consistent Born approximation (SCBA). Furthermore, a procedure based on EHT basis set modification is described. We use this model to study the effect of the temperature dependent dephasing, due to low lying modes in far-infrared range for which $\hbar\omega\ll k_BT$, on the \textit{resonant} conduction through highest occupied molecular orbital (HOMO) level of a phenyl dithiol molecule sandwiched between two fcc-Au(111) contacts. Furthermore, we propose to include dephasing in room temperature molecular resonant conduction calculations. 
\end{abstract}

\maketitle
\section{Introduction} We present an extended H\"uckel theory (EHT) based self-consistent mean-field transport model for incoherent single molecule spectroscopy. Although simple, this approach is worth-pursuing because EHT is numerically inexpensive and overcomes many of the short-comings of the more sophisticated theories \cite{Raza07} and has been used for nanoelectronics \cite{Raza08}. Furthermore, the simplicity and the exponential dependence of the Slater type orbitals basis set allows an intuitive solution to tip modeling as described in this paper. The proper tip modeling is important because if condensed matter like short ranged orbitals are used, one has to bring the tip very close to the molecule [about 4-5$\AA$] to obtain the experimentally observed current levels. This artificially results in higher than the actual voltage drop across the molecule resulting in not only wrong Laplace's potential but Hartree potential is also overestimated. 

\begin{figure}
\vspace{2.8in}
\hskip1.5in\includegraphics{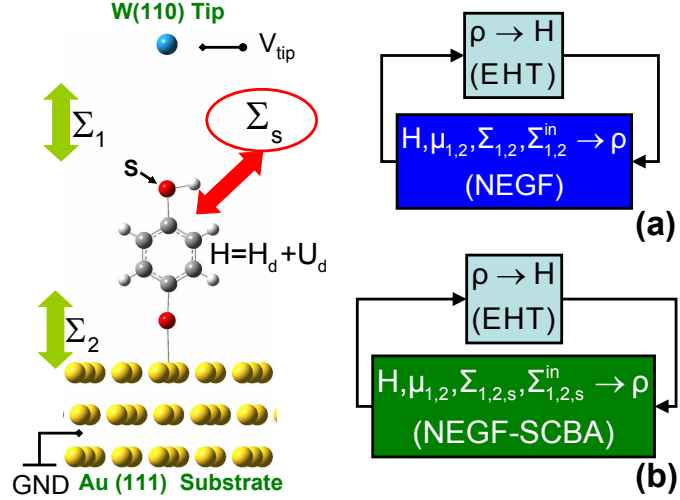}
\caption{(color online) Schematic diagram. Scanning Tunneling Spectroscopy (STS) setup for probing conduction through phenyl/benzene dithiol (PDT/BDT) molecule bonded to Au(111) substrate. Voltage is applied at the tungsten W(110) tip. (a) A schematic showing the self-consistent procedure of solving EHT-NEGF equations for coherent transport \cite{Ferdows05}. (b) A schematic detailing the self-consistent procedure of solving EHT-NEGF-SCBA equations for quantum transport with elastic and inelastic dephasing.}
\end{figure}

Furthermore, the electron-phonon scattering in molecular structures on different metallic substrates has been studied quite extensively at low temperature experimentally \cite{Hu98,Reed04,Shashidhar04,Wanga06} and theoretically \cite{Jauho04,Guo05,Ventra05,Magnus06,Ratner05, Ratner06, Kawai04} in the form of off-resonant inelastic tunneling spectroscopy (IETS). In IETS, the molecule acts as the scattering region and the distinct peaks in the second derivative of I-V characteristics identify the molecular vibrations. Moreover, in electron-phonon scattering, not only electron exchanges energy with the phonon bath but also its phase is irreversibly lost due to phonon degree of freedom. If the phonon energy ($\hbar\omega$) is less than a characteristic energy (\textit{e.g.} $k_BT$), this inelastic dephasing process may be approximated by an elastic dephasing event for numerical convenience. Since all the energy channels are coupled in inelastic dephasing, solving for it is computationally challenging. 

Apart from this, the phonon modes of a molecule bonded to a substrate are different from gas phase. Organic molecules in gas phase have vibrational energies in the infrared range, \textit{i.e.} 100-350meV. However, experiments using IETS \cite{Shashidhar04}, high resolution electron energy loss spectroscopy (HREELS) \cite{Kawai04} and low temperature transport \cite{Park00} suggest that bonding to a substrate leads to low lying modes in the far-infrared range. Dominant modes having vibrational energies less than about 10meV have been reported theoretically \cite{Guo05, Ventra05}. These far-infrared modes are the result of the combined motion of the molecule with respect to the contacts. Most of these infra-red modes are longitudinal modes and couple well with carriers \cite{Guo05, Ventra05}. Hence, the low lying modes, for which $\hbar\omega\ll k_BT$ at room temperature, are the most important phonon modes for molecules bonded to substrates. In a previous study \cite{Raza06}, we report transport calculations at room temperature for a styrene chain bonded to $n^{++}$-H:Si(001) substrate. In Ref. \cite{Raza06}, we propose to include dephasing due to low lying modes in order to theoretically reproduce experimental results.

Furthermore, in context of molecular spectroscopy, there are four sources of broadening: (1) contact broadening which is about 0.1-0.3eV for a good contact (2) the Fermi function broadening of the two contacts (3) broadening due to Hartree self-consistency which is usually small in STS setup due to tip being about 1nm away from substrate and (4) broadening due to electron-phonon scattering which is temperature dependent due to phonon population given by Bose-Einstein factor as $1/[exp(\hbar\omega / k_BT)-1]$. The main focus of this work is the broadening due to this fourth source in transport through a phenyl/benzene dithiol (PDT/BDT) molecule bonded to Au(111) substrate as shown in Fig. 1. The tip configuration used is tungsten W(110) having a workfunction of about 5.2eV. 

The paper is divided into five sections. In Sec. II, we describe the mean-field self-consistent EHT-NEGF-SCBA model for studying transport with dephasing due to electron-phonon scattering. Furthermore, details are provided about the electrostatic calculation. Tip modeling is an important aspect of the calculations and is discussed in Sec. III. In Sec IV, we discuss the results and report the temperature dependence of the transport quantities on the electron-phonon scattering due to low lying modes. Finally, we provide the conclusions in Sec. V. The molecular visualization is done using GaussView \cite{GW03}. 

\begin{figure}
\vspace{3in}
\hskip1.5in\includegraphics{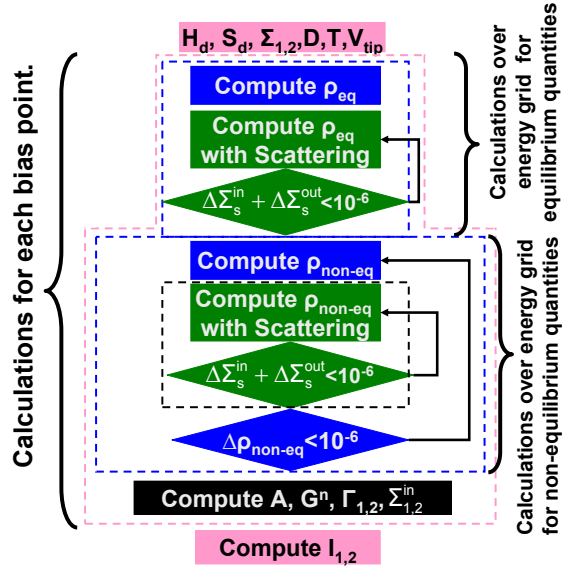}
\caption{(color online) Block diagram for transport with elastic dephasing. For elastic dephasing, all the energy channels are independent. Therefore, energy dependent quantities can be calculated independently for each energy point with the convergence criteria shown. Equilibrium and non-equilibrium quantities for each bias point are calculated as shown. The green and blue blocks are color coded to match with the blocks in Fig. 1(a,b). For inelastic dephasing, since energy channels are coupled, one has to use convergence criteria on the calculated energy dependent quantities over the whole energy range.}
\end{figure}

\section{Formalism} We use the single particle non-equilibrium Green's function formalism (NEGF) in the mean field approximation using non-orthogonal basis to model the quantum transport. We define the time-retarted non-equilibrium Green's function in non-orthogonal basis as \cite{Datta05,Ferdows03}:
\begin{eqnarray}G(E,V_{tip})=[(E+i0^+)S-H-E_{fc}S-\Sigma_{1,2}-\Sigma_{s}]^{-1}\end{eqnarray}
where S is the overlap matrix and $H=H_d+U_d$. $H_d$ is the device Hamiltonian for an isolated molecule expressed in EHT \cite{Hoffmann}. $\Sigma_{1,2}$ are the self-energies of W(110) tip and Au(111) contact respectively - calculated recursively using EHT as well. Since, there is no vacuum reference in EHT parametrization, we use input from experiments or other sophisticated theoretical methods to set the reference. $E_{fc}$ is used as a parameter to achieve this objective. It is taken 2eV in this paper to match the experimentally observed peak in conductance spectrum. $U_d$ incorporates both Laplace's potential due to the applied tip voltage ($V_{tip}$) and Hartree potential due to the non-equilibrium density of electrons in the molecular channel. The contact self-energies $\Sigma_{1,2}$ are related to the broadening functions as $\Gamma_{1,2}=i(\Sigma_{1,2}-\Sigma_{1,2}^\dag)$. The tip self-energy ($\Sigma_{1}$) is further discussed in Sec. III. The Au(111) self-energy ($\Sigma_{2}$) has been discussed in Ref. \cite{Ferdows05}. The spectral function is defined as $A=i(G-G^\dag)$, which is related to the density of states of the molecular region as $D(E)=2(for\ spin)\times tr(AS)/2\pi$. The electron correlation function $G^n (=-iG^<)$ is given as:
\begin{eqnarray}G^n=G(\Sigma^{in}_{1}+\Sigma^{in}_{2}+\Sigma^{in}_{s})G^\dag\end{eqnarray}
where $\Sigma^{in}_{1,2}(=-i\Sigma^{<}_{1,2})$ are the contact inflow functions defined as $\Gamma_{1,2}f_{1,2}$ and $\Sigma^{in}_{s}(=-i\Sigma^{<}_{s})$ is the scattering inflow function. $f_{1,2}(E)$ are the contact Fermi functions given as $1/[1-exp((E-\mu_{1,2}) / k_BT)]$. $\mu_{1,2}$ are the chemical potentials of the two contacts given as $\mu_1=\mu_{o}-eV_{tip}$ and $\mu_2=\mu_{o}$ respectively - $\mu_{o}$ is the equilibrium chemical potential. In this model, contacts are assumed to remain in equilibrium. This is a valid assumption for metallic contacts and degenerately doped semi-conducting contacts. However for the moderately doped or lightly doped semi-conducting contacts, band bending due to $V_{tip}$ makes these calculations computationally expensive. The hole correlation function is defined as $G^p=A-G^n$. The electron density matrix is then given as:
\begin{eqnarray}\rho=\frac{1}{2\pi}\int_{-\infty}^{\infty}dE\ G^n(E)\end{eqnarray}
The total number of electrons is computed as $N=2(for\ spin)\times tr(\rho S)$.
Finally, the I-V characteristics are computed as follows \cite{Datta05,Meir92}: 
\begin{eqnarray}{I_i}(V_{tip})=2(for\ spin)\times\frac{e}{h}\int_{-\infty}^{\infty}dE\ tr(\Sigma_i^{in}A-\Gamma_iG^n)\end{eqnarray}
Since scattering self-energy ($\Sigma_s$), G, $G^n$ and $\Sigma^{in}_s$ depend on each other, we solve these four quantities self-consistently along with the Hartree self-consistent loop as shown in Fig. 1(b). The flow diagram for the self-consistent procedure at each bias is shown in Fig. 2. For SCBA, it can be shown that the current through the \textit{scattering contact} is always zero.

The scattering inflow function ($\Sigma^{in}_s$), outflow function ($\Sigma^{out}_s$) and broadening function ($\Gamma_s$) are defined as \cite{Mahan87}:

\begin{eqnarray}\Sigma^{in}_s(E)=\int_0^{\infty}\frac{d(\hbar\omega)}{2\pi}D^{em}(\hbar\omega)SG^n(E+\hbar\omega)S+D^{ab}(\hbar\omega)SG^n(E-\hbar\omega)S\end{eqnarray}
\begin{eqnarray}\Sigma^{out}_s(E)=\int_0^{\infty}\frac{d(\hbar\omega)}{2\pi}D^{em}(\hbar\omega)SG^p(E-\hbar\omega)S+D^{ab}(\hbar\omega)SG^p(E+\hbar\omega)S\end{eqnarray}
\begin{eqnarray}\Gamma_s(E)=\Sigma^{in}_s+\Sigma^{out}_s\\
=\int_0^{\infty}\frac{d(\hbar\omega)}{2\pi}D^{em}(\hbar\omega)S[G^n(E+\hbar\omega+G^p(E-\hbar\omega)]S\\+D^{ab}(\hbar\omega)S[G^n(E-\hbar\omega)+G^p(E+\hbar\omega)]S\end{eqnarray}
where $D^{em}(\hbar\omega)=(N+1)D_o(\hbar\omega)$ and $D^{ab}(\hbar\omega)=N\ D_o(\hbar\omega)$ are the emission and absorption dephasing functions respectively. N is the equilibrium number of phonons given by Bose-Einstein statistics as $1/[exp(\hbar\omega / k_BT)-1]$. Eq. 5 implies that electron inflow at a particular energy E is dependent on the electron correlation function $G^n$ at energy $E+\hbar\omega$ for emission and $E-\hbar\omega$ for absorption. This relationship is reversed for the electron outflow as in Eq. 6. The overall broadening due to this scattering process is given by $\Gamma_s(E)$. Furthermore, $D_o(\hbar\omega)$ is a fourth ranked tensor and is related to the electron phonon interaction potential (U) as $<i,j|U^\dagger U|k,l>$. Thus, $D_o(\hbar\omega)$ can be different for different molecular orbitals and is explicitly written as $D_o(i,j;k,l;\hbar\omega)$. Eqs. 5-7 are modified from those reported in Ref. \cite{Mahan87} for non-orthogonal basis as discussed in Ref. \cite{Ferdows03}.

\textit{High energy phonon limit:} For $\hbar\omega \gg k_BT$, $N\approx 0$ and $N+1\approx 1$, hence $D^{em}\approx D_o$ and $D^{ab}\approx 0$. Thus, Eq. 7 becomes:
\begin{eqnarray}\Gamma_s(E)=\int_0^{\infty}\frac{d(\hbar\omega)}{2\pi}D_o(\hbar\omega)S[G^n(E+\hbar\omega)+G^p(E-\hbar\omega)]S\end{eqnarray}
Thus, contribution due to absorption is negligible in this case because not enough phonons are available to be absorbed. Moreover, since $\Gamma_s(E)$ is temperature independent, the broadening due to electron-phonon scattering for $\hbar\omega \gg k_BT$ is approximately temperature independent.

\textit{Low energy phonon limit:} For $\hbar\omega\ll k_BT$, by keeping the first term in Taylor series expansion of $exp(\hbar\omega\ k_BT) \approx 1+\hbar\omega // k_BT$, one obtains $N+1 \approx N \approx {k_BT}/{\hbar\omega}$. This leads to,
\begin{eqnarray}\Gamma_s(E)=\int_0^{\infty}\frac{d(\hbar\omega)}{2\pi}[D^{em}(\hbar\omega)+D^{ab}(\hbar\omega)]SA(E)S\end{eqnarray}
and 
\begin{eqnarray}D^{em}(\hbar\omega)\approx D^{ab}(\hbar\omega)=N\ D_o(\hbar\omega)\approx \frac{k_BT}{\hbar\omega} D_o(\hbar\omega)\end{eqnarray}
We finally obtain $\Gamma_s(E)$ as,
\begin{eqnarray}\Gamma_s(E) \approx \underbrace{T\int_0^{\infty}\frac{d(\hbar\omega)}{2\pi}\frac{2k_BD_o(\hbar\omega)}{\hbar\omega}}_{D}SA(E)S\end{eqnarray}
where D is referred to as dephasing strength. As shown above, the broadening due to elastic dephasing is directly proportional to temperature. 

Furthermore, the real part of $\Sigma_s(E)$ is computed by taking a Hilbert transform of the imaginary part as below, \begin{eqnarray}\Sigma_s(E)=\frac{1}{\pi}\int_{-\infty}^{\infty}dy\frac{-\Gamma_s(y)/2}{E-y}-i\Gamma_s(E)/2 \end{eqnarray}
Calculating the real part of $\Sigma_s(E)$ is computationally expensive and leads to convergence issues. Previously, it has been reported \cite{Magnus06} that this real part contributes on the order of meV. We therefore ignore it in this paper. We phenomenologically approximate the dephasing strength (D) by a constant. This approximation treats $T_1$ and $T_2$ time in an average manner and seems to reproduce experimental results \cite{Raza06}. Treating D as a constant is similar to the dephasing strength used in a B\"uttiker probe. However, the mathematical framework of SCBA is of course different from that of B\"uttiker probe method. Furthermore, a rigorous calculation of dephasing functions and dephasing strength can be readily incorporated in our model. Apart from this, SCBA has some limitations, \textit{e.g.} it does not take care of multiple phonon processes implicitly for a single scattering event and it does not include crossed diagrams which correspond to successive emission and re-absorption of phonons by the same electron. 

\begin{figure}
\vspace{4in}
\hskip2in\includegraphics{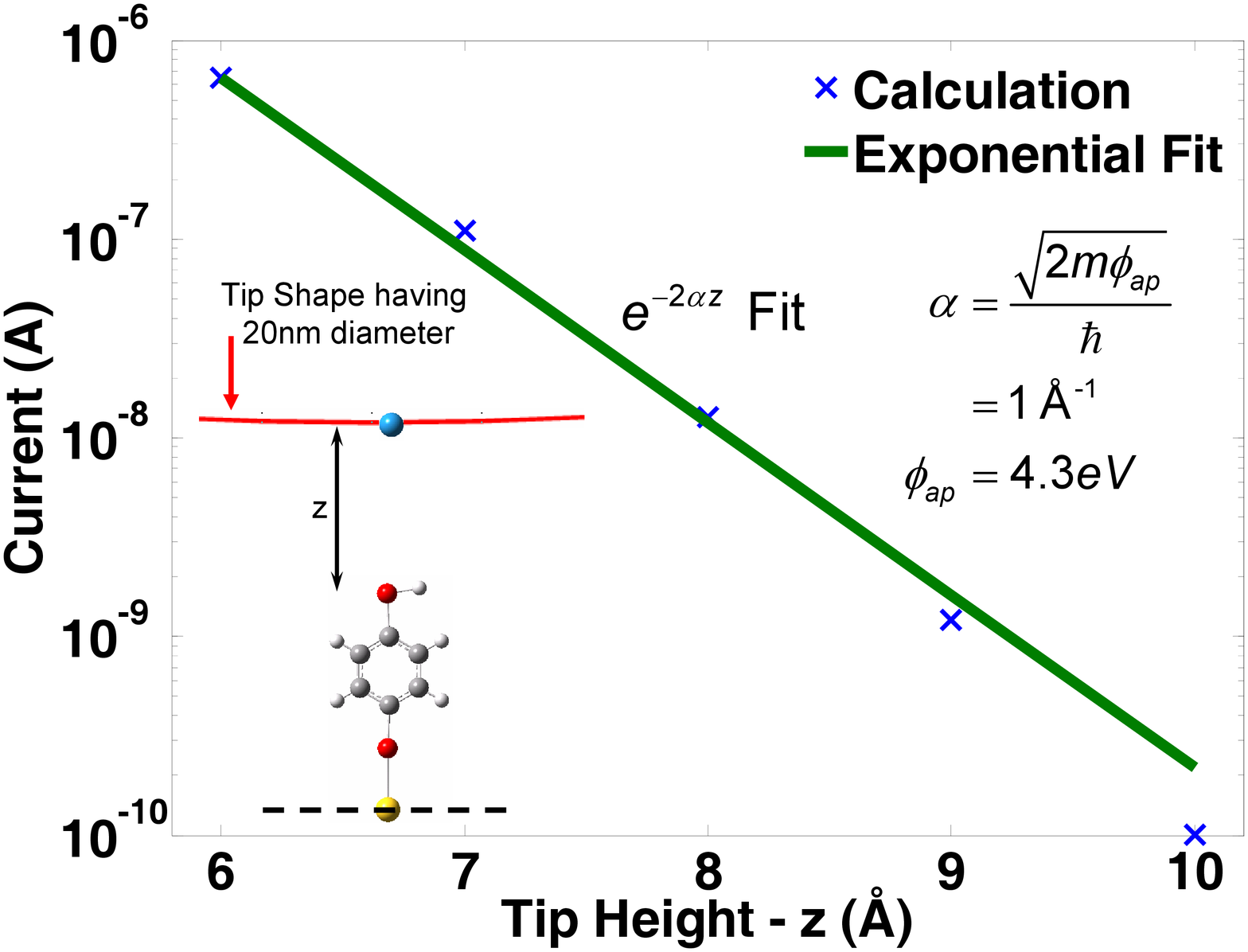}
\caption{(color online) Tip modeling. The calculated current (I) for $V_{tip}=1V$ as a function of tip-molecule distance (z) showing exponential dependence of current on z. The extracted barrier height from this I-z plot is 4.3eV, whereas with original parameters it is approximately 10eV. The calculated apparent barrier height ($\phi_{ap}$) is close to the work functions of the contacts being used. Furthermore for calculating Laplace and image potential due to $V_{tip}$, a 20nm diameter tip is used as shown in the inset. The tip is $7\AA$ away from the molecule in this visualization.}
\end{figure}

We use a finite-element solver \cite{COMSOL} for solving Laplace's equation in three dimension to obtain potential drop across molecule due to $V_{tip}$. This method also provides the zero bias band line up potential due to the Fermi level mismatch of 0.1eV between Au(111) and W(110) tip, whose work functions are 5.1eV and 5.2eV respectively. We take proper tip shape into account. A 20nm tip diameter is assumed as show in the inset of Fig. 3. The Hartree potential for the molecule is approximated via the complete neglect of differential overlap method \cite{Ferdows05,Pople}. Image effects are incorporated to ensure that self-consistency is not over-estimated. 

\section{Tip Modeling} For a good tip which can give atomic resolution, the last atom at tip apex dominates STS. Therefore, we calculate the overlap between molecule and tip for this single tungsten atom. This assumption is for a "good" tip without any adsorbate or multiple tip structures which may give ghost images in scanning tunneling microscopy (STM). We use W(110) configuration since it is the lowest energy state. Furthermore for working tip heights, electronic effects due to the wave-function overlap between the tip atoms and the molecule become important. This overlap gives exponential dependence of the decay of tunnel current (I) with tip height (z). Within WKB approximation, apparent barrier height ($\phi_{ap}$) can be calculated from the slope of the \textit{log}(I)-z plot. This apparent barrier height is usually found to be close to the workfunctions of the materials being used - Au(111) in our case. We find that the apparent barrier height is about 10eV with unmodified EHT parameters, which is unphysical. We report a scheme for modifying basis set of tip atom to get correct apparent barrier height. One does not need to modify the basis set of atoms in the molecule because they are already optimized for gas phase. Similar approach has been used before in the context of scanning tunneling microscopy  \cite{Cerda97}. With the modified EHT parameters of the s-orbital basis set \cite{TipParameters} for tungsten atom, we obtain $\phi_{ap}=4.3eV$ from the calculated I-z plot with $V_{tip}=1V$ as shown in Fig. 3.  

\begin{figure}
\vspace{2.9in}
\hskip1.5in\includegraphics{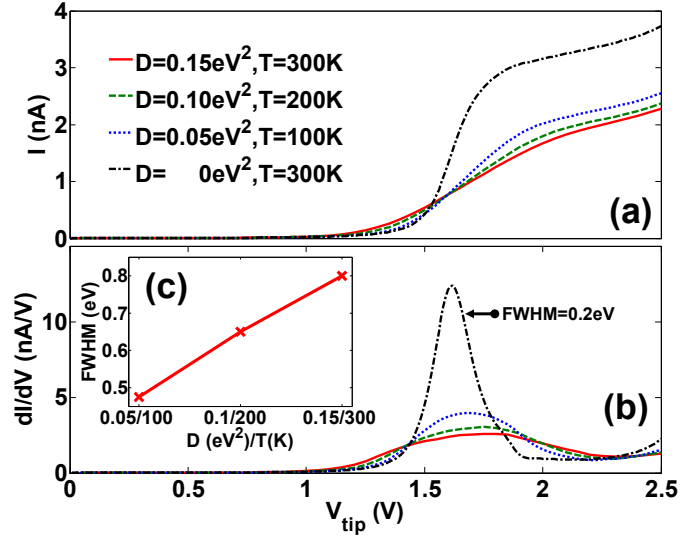}
\caption{(color online) Incoherent resonant conduction through the highest occupied molecular orbital (HOMO) level with elastic dephasing due to low energy electron-phonon scattering. The dephasing strength (D) is directly proportional to the temperature (T). Thus, the additional broadening due to elastic dephasing decreases with decreasing temperature in a linear fashion. (a) I-V characteristics showing broadening of spectroscopic features due to dephasing. (b) dI/dV-V characteristics providing an alternate and more detailed view of broadening of spectroscopic features. (c) Full width half maximum (FWHM) of the first conductance peak corresponding to the HOMO level showing linear dependence on temperature. The dephasing strength (D) is $0.15eV^2$ at 300K to give experimentally observed broadening.}
\end{figure}

The above parameter modification affects $S_1$ and $H_1$, provided the tip self-energy $\Sigma_{1}$ is defined as $[(E+i0^+)S_1-H_1]g_{s1}((E+i0^+)S_1^\dag-H_1^\dag)$, where $g_{s1}$ is the surface Green's function for the tip.  We assume a constant $g_{s1}$, calculated from the density of states at equilibrium chemical potential of bulk tungsten $D_W(\mu_o)$, \textit{i.e.} $g_{s1} = -i\pi D_W(\mu_o)$ - a commonly used approximation [see \textit{e.g.} Ref. \cite{Raza06} and references there-in]. However, $\Sigma_{1}$ is still energy dependent and this energy dependence should in fact capture all the barrier tunneling effects. However, it would not be able to capture the features due to peculiar electronic structure of tip as in Ref. \cite{Datta99}. 

\section{Results} In this section, we apply the proposed EHT-NEGF-SCBA model to incoherent transport with elastic dephasing through the highest occupied molecular orbital (HOMO) level of a PDT/BDT molecule bonded to a Au(111) substrate through a single Au-S bond and probed using a tungsten W(110) tip as shown in Fig. 1. Au-S bond length is the standard $2.53\AA$ and tip molecule distance is $10\AA$. The dephasing strength (D) at 300K is taken as $0.15eV^2$ to obtain experimentally observed broadening. Dephasing strength at any other temperature T is then calculated following Eq. 11. 

The calculated I-V characteristics are shown in Fig. 4(a). For coherent conduction through PDT/BDT molecule at 300K (D=0), the I-V curve has a sharp rise. Fig. 4(b) provides an alternate view showing the dI/dV-V characteristics. The full width half maximum (FWHM) of the conductance peak is about 0.2eV. This broadening is due to (1) the finite life-time of electrons in the molecule as a result of the coupling with the contacts given by the broadening functions $\Gamma_{1,2}$, (2) contact Fermi functions giving about $5.4k_BT$ broadening, and (3) Hartree potential, which is small because tip is 10$\AA$ from molecule. 

With $D=0.15eV^2$ at 300K, the I-V curve gets smeared out due to spectral broadening induced by elastic dephasing and overall current level decreases as shown in Fig. 4(a). Furthermore, current starts increasing earlier and hence is slightly higher than D=0 case for $V_{tip} < 1.45V$. Correspondingly, the dI/dV plot [see Fig. 4(b)] has a higher conductance value for $1.45V > V_{tip} > 1.8V$, (\textit{i.e.} before and after the conductance peak). The FWHM of this conductance peak is about 0.8eV as shown in Fig. 4(c). The additional 0.6eV broadening is due to elastic dephasing. Such a high broadening has been observed experimentally in Ref. \cite{Tian98}. Apart from this, the comparison of I-V and dI/dV-V plots for D=0 and $D=0.15eV^2$ at 300K raises two important questions (1) why the overall current value decreases and (2) how does the temperature dependence of dephasing strength affect the FWHM of conductance peaks. 

The answer to the first question depends on the energy dependence of the contact broadening functions $\Gamma_{1,2}$. After including dephasing, the spectral weight increases in the energy range where it was smaller before and \textit{vice versa}. If the energy region, where the spectral weight is transferred, has larger broadening functions $\Gamma_{1,2}$, it results in a higher current. We analyze the energy dependence of $\Gamma_{1,2}$ and find that $\Gamma_{1,2}$ decrease as a function of decreasing energy, which corresponds to positive $V_{tip}$. Thus, the broadened molecular spectral function gives rise to higher current values for lower $V_{tip}$, where the broadening functions $\Gamma_{1,2}$ are larger. Similarly, current values are smaller for larger $V_{tip}$, where $\Gamma_{1,2}$ are smaller. Hence, the overall current level decreases with dephasing. However, in a previous study \cite{Raza06} with $n^{++}-H:Si(001)-(2\times 1)$ contact, we conclude that the overall current level increases with dephasing. In Ref. \cite{Raza06}, the HOMO level is near valence band-edge, where the contact density of states (DOS) is small resulting in a small current. After including dephasing, the HOMO level gets broadened and a portion of it is in energy range where the DOS is large. Thus, dephasing in this case results in an order of magnitude higher current. 

\begin{figure}
\vspace{2.75in}
\hskip1.5in\includegraphics{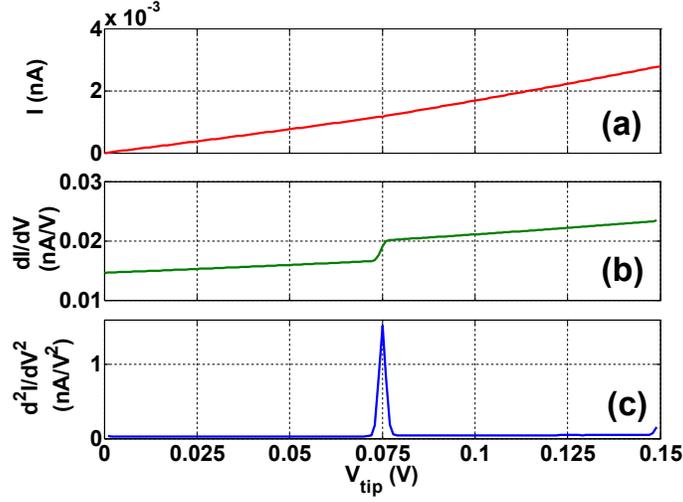}
\caption{(color online) Effect of inelastic scattering on the off-resonant conduction at 4K assuming one phonon mode with $D^{em}=0.1eV^2$ and $\hbar\omega=75meV$. (a) IETS signature is not noticeable in the I-V characteristics. (b) Inelastic scattering results in a step in dI/dV-V characteristics. (c) IETS spectra ($d^2I/dV^2$-V characteristics) showing a peak at 75meV corresponding to the phonon mode. Since $\hbar\omega \gg k_BT$, $D^{ab}=D^{em}e^{-\hbar\omega/k_BT} \approx 0$. Thus, the IETS peak is due to emission of phonons and processes due to absorption of phonons are weak.}
\end{figure}

With decreasing temperature and dephasing strength, \textit{e.g.} at 200K and 100K with $D=0.1eV^2$ and $D=0.05eV^2$ respectively, the I-V plots follow same trend as shown in Fig. 4(a,b). The calculated FWHM values of these conductance peaks are shown in Fig. 4(c) to quantify the effect of dephasing. FWHM is directly proportional to the temperature and dephasing strength. The dependence of FWHM of conductance peak on temperature and dephasing strength could become complex if temperature dependent structural changes are present.

In Fig. 5, we report the inelastic tunneling spectroscopy (IETS) of PDT/BDT molecule at 4K. The purpose of these standard and well-established results is to convey the message that the model is able to handle calculations for inelastic dephasing as well. We assume a single phonon mode with energy $\hbar\omega=75meV$ and emission dephasing function $D^{em}=0.1eV^2$. The absorption dephasing function is then given as $D^{ab}=D^{em}e^{-\hbar\omega/k_BT}$. Fig. 5(a) shows the I-V characteristics with no feature of IETS signal. Fig. 5(b) shows the dI/dV-V characteristics with a step in the conductance curve due to emission of phonons at $eV_{tip}=\hbar\omega=75meV$. The same signal appears as a peak in the $d^2I/dV^2$ spectrum as shown in Fig. 5(c). Furthermore, we include the Hartree self-consistency for elastic dephasing only. Since, we are studying the inelastic dephasing in the off-resonance regime, self-consistency is not important. 
 
\section{Conclusions} We have reported an EHT-NEGF-SCBA model and have used it to study elastic and inelastic dephasing. It has been previously reported that low lying modes are important in molecular conduction at low temperature \cite{Shashidhar04,Kawai04,Guo05,Ventra05}. We suggest that they should be important at room temperatures as well and should be included in transport calculations. Furthermore for $\hbar \omega \ll k_BT$, inelastic dephasing can be approximated by elastic dephasing. Within this approximation, it can be shown that the broadening is proportional to temperature. Moreover, for Au contacts, the overall current level tends to reduce due to this dephasing. This could help bridge the gap between theory and experiment for resonant conduction along with other proposals \cite{Kirczenow01, Bhaskaran06}, where theoretically calculated currents are higher as compared to the experimental observations \cite{Ventra00}. Finally, we present results for IETS to show that within the model, inelastic dephasing can be handled as well. 

It is a pleasure to acknowledge useful discussions with S. Datta and E. C. Kan. We thank F. Zahid and T. Raza for H\"uckel-IV 3.0 \cite{Ferdows05} codes and later for reviewing the manuscript. Computational facilities were provided by the NSF Network for Computational Nanotechnology.


\begin{thebibliography}{10}
\bibitem{COMSOL} COMSOL is a trademark of COMSOL AB (comsol.com).
\end{thebibliography}
\end{document}